\pdfoutput=1
\documentclass[twocolumn,showpacs,prb,amsmath,amssymb,floatfix]{revtex4-1}

\usepackage{amsmath,amssymb}
\usepackage{bm}
\usepackage{epsfig}
\usepackage{psfrag}
\usepackage[svgnames]{xcolor}

\newcommand{\bd}{\bm}

\begin{document}

\title{Quantum criticality of dipolar spin chains}

\author{Aldo Isidori, Annika Ruppel, Andreas Kreisel, and Peter Kopietz}

\affiliation{Institut f\"{u}r Theoretische Physik, Goethe-Universit\"{a}t  Frankfurt,  
60438 Frankfurt, Germany}

\author{Alexander Mai and Reinhard M. Noack}

\affiliation{Fachbereich Physik, Philipps-Universit\"{a}t Marburg, 35032 Marburg, Germany}

\date{September 23, 2011}

 \begin{abstract}
We show that a chain of Heisenberg spins interacting with 
long-range dipolar forces in a magnetic field $h$ perpendicular
to the chain
exhibits a quantum critical  point belonging
to the two-dimensional Ising universality class.
Within linear spin-wave theory 
the magnon dispersion for small momenta $k$
is $ [\Delta^2 + v_k^2 k^2 ]^{1/2}$, where
$\Delta^2 \propto | h - h_c|$ and $v_k^2 \propto |\ln k | $.
For fields close to $h_c$ linear spin-wave theory breaks down
and we investigate the system using 
density-matrix and functional renormalization group methods. 
The Ginzburg regime where non-Gaussian 
fluctuations are important is found to be 
rather narrow on the ordered side of the transition, and
very broad on the disordered side.

\end{abstract}


\maketitle

\section{Introduction}

The  long-range nature and spatial anisotropy of the
dipole-dipole interaction in quantum many-body systems
can give rise to unconventional effects 
such as exotic ordered phases and excitation spectra.\cite{Baranov08}
The experimental study of these phases is now possible
due to substantial progress in controlling
the parameters of trapped ultracold atoms.
While both bosonic\cite{Stuhler05,Lahaye07,Lu11} and 
fermionic\cite{Ni08} dipolar quantum gases have been
realized, it remains a challenge to
design purely dipolar spin systems by localizing ultracold
atoms or molecules with permanent magnetic or electric moments
on an optical lattice.
A promising strategy to obtain experimental realizations of
dipolar magnets with localized spins 
uses trapped ions, which were recently employed to
design spin Hamiltonians with controllable interactions between the 
spins.\cite{Friedenauer08,Kim10}

Heisenberg magnets with dipole-dipole interactions
in two and three dimensions
have been investigated theoretically for more than half a 
century,\cite{Akhiezer46,Maleev76,Syromyatnikov06}
but one-dimensional dipolar spin chains
have not received much attention. 
This may be
due the fact that in condensed matter systems the
exchange interaction is usually much larger than the dipole-dipole
interaction, so that it has not been possible to realize experimentally 
purely dipolar spin chains before the emergence of the field of ultracold atoms.
Recently several authors\cite{Kestner11,Gorshkov11,DallaTorre06}
pointed out that tunable spin chains with dipole-dipole interactions 
can be derived from  two-component dipolar gases
as effective models for the spin degrees of freedom.
The investigations presented in this work are motivated by the expectation that
in the near future it will be possible
to design purely dipolar spin chains using trapped atoms or ions
at ultra-low temperatures. 
 
In a previous study of dipolar spin chains\cite{Singh86}
the long-range dipole-dipole interaction
was truncated at the next-nearest neighbor, 
which misses the logarithmic correction to the spin-wave velocity discussed below.
Moreover, the spin-wave calculations of Ref.~\onlinecite{Singh86} did
not take into account the tilted geometry of
the classical ground state in the low-field phase,
thus missing the quantum critical point which separates the
tilted phase from the high-field phase where all spins
align with the magnetic field.

\section{Classical ground state and spin-wave expansion}

\begin{figure}[tb]    
   \centering
  \includegraphics[width=0.22\textwidth]{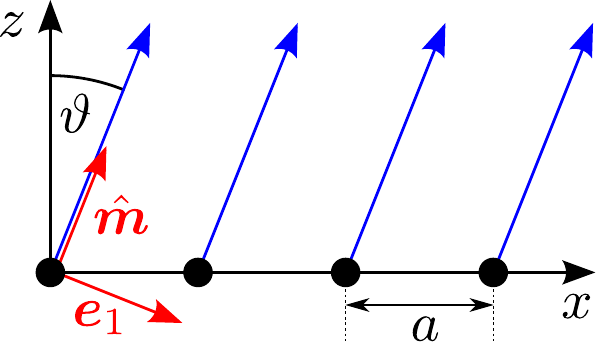}
  \caption{%
(Color online)  Classical ground state of the
dipolar spin chain defined in Eq.~(\ref{eq:Hamiltonian2}) for $h < h_c$.
Big arrows represent the vectors $S \hat{\bd{m}}$ at the
lattice sites (black dots).
Small arrows represent the tilted basis vectors
$\hat{\bd{m}}$ and $ \bd{e}_1$.
The third basis vector $\bd{e}_2 = \hat{\bd{y}}$ points into the plane of the
paper.
}
  \label{fig:tilt}
  \end{figure}

We consider a chain of quantum spins  $\bd{S}_i$ of 
spin $S$
in an external magnetic field $ \bd{h}$ perpendicular to the chain 
which are coupled by both dipolar and exchange interactions.
Choosing our coordinate system such that
the chain lies along the $x$-axis and 
the external magnetic field $ \bd{h} = h \hat{\bd{z}}$  points in the
$z$-direction, our Hamiltonian reads
\begin{eqnarray}
 {\cal{H}}   & = & - \frac{1}{2} \sum_{ i j, i \neq j } 
 \frac{ \mu^2 }{ | x_{i} - x_{j} |^3}
 \left[  3  {{S}}_i^x   
  {{S}}_j^x   -  {\bd{S}}_i \cdot {\bd{S}}_j   \right]
 \nonumber 
 \\
&  & 
 - \frac{1}{2} \sum_{ i j  } J_{ij} {\bd{S}}_i \cdot {\bd{S}}_j
  -     h \sum_{i} {{S}}^z_i,
 \label{eq:Hamiltonian2}
 \end{eqnarray}
where sums are over the $N$ sites $x_i$  of a one-dimensional 
lattice with spacing $a$.
The long-range dipolar  interaction is characterized by an effective 
magnetic moment
$\mu = g \mu_B $, where $g$ is the effective gyromagnetic factor 
and $\mu_B $ is the Bohr magneton.
We assume that the spins are also coupled by
nearest-neighbor ferromagnetic exchange interactions, i.e.,
$J_{ij} = J >0$ if $ | x_i - x_j| = a $ and $J_{ij} =0$ otherwise.
Due to the competition between the Zeeman energy, which favors 
alignment of the spins along the $z$-axis, and the
dipolar interaction, which favors spin alignment along the $x$-axis, 
the spins align with a finite tilt angle $\vartheta$ relative to the
field direction below a certain critical field $h_c$, as shown in Fig.~\ref{fig:tilt}.
The magnetization points then in the direction of the unit vector
$ \hat{\bd{m}} = \sin \vartheta \hat{\bd{x}} + \cos \vartheta \hat{\bd{z}}$.
The tilt angle $\vartheta$ in the classical ground state ($S \rightarrow \infty$)
can be determined by replacing the
spin operators $\bd{S}_i$  
in Eq.~(\ref{eq:Hamiltonian2}) by classical vectors
$S \hat{\bd{m}}$ of length $S$, which yields the classical ground state energy per site
 \begin{equation}
 {\cal{H}}_0 / N  = -  J S^2 - h S \cos \vartheta 
- D_0 S ( 3 \sin^2 \vartheta -1 )/6,
 \label{eq:H0}
 \end{equation}
where 
$D_0$ is controlled by the dipole-dipole interaction,
 \begin{equation}
 D_0 = \frac{3S}{N} \sum_{i j, i \neq j } \frac{  \mu^2}{ | x_{i} - x_{j} |^3 }
 = 6 \zeta ( 3 ) \frac{ S \mu^2 }{a^3}.
 \label{eq:hc}
 \end{equation}
Minimizing  ${\cal{H}}_0$ with respect to the tilt angle $\vartheta$ we find
$\cos \vartheta  = h/D_0$ for $ h \leq D_0$,
and  $\vartheta =0$ for $ h > D_0$.
In this work we show that 
at the critical  magnetic field $h_c$ where the tilt angle $\vartheta$ vanishes
the system undergoes a continuous quantum phase transition.
Classically the critical field is $h_c= D_0$, but
for small $S$ the value of $h_c$ is
substantially smaller than $D_0$ due to  quantum fluctuations.
Since Hamiltonian (\ref{eq:Hamiltonian2}) 
has no continuous spin symmetry, and
the $Z_2$ symmetry
$S^x_i \rightarrow - S^x_i$
is spontaneously broken in the tilted phase, we expect that the 
quantum phase transition belongs to the universality class
of the two-dimensional Ising model.

\begin{figure*}[tb]    
   \centering
\includegraphics[width=.8\linewidth]{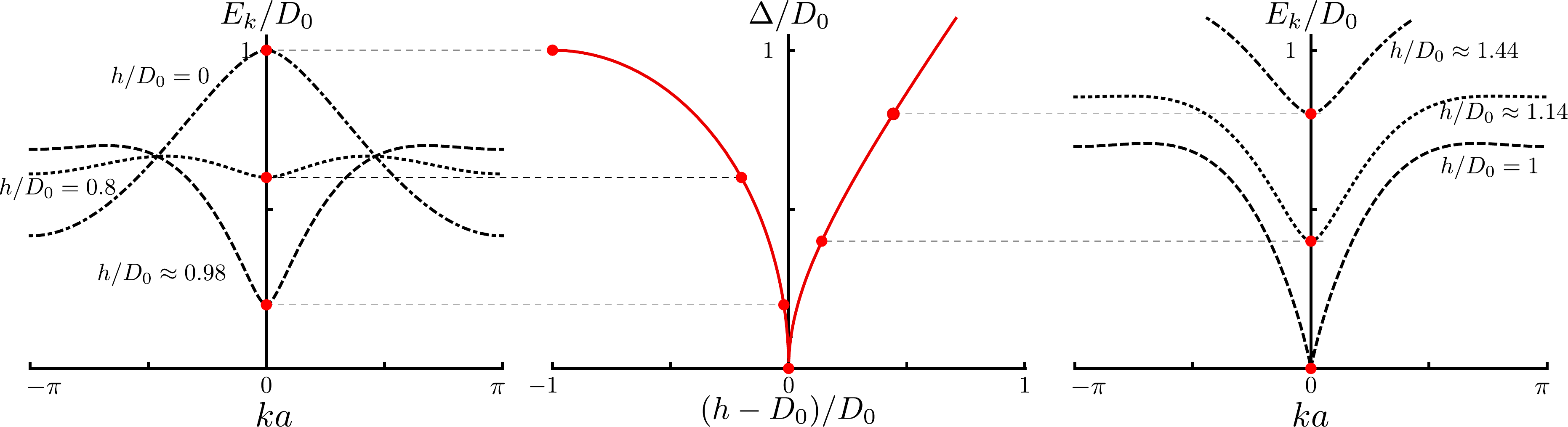}
  \caption{%
(Color online) Graph of the spin-wave dispersion $E_k$
for $J=0$ and  $h < D_0$ (left), and for $h > D_0$ (right). 
In the middle we show the gap $\Delta$ as a function of $(h-D_0)/D_0$.
}
    \label{fig:dispersion}
  \end{figure*}

To calculate the spin-wave spectrum, we expand the
spin operators in the  tilted basis $\{ \bd{e}_1, \bd{e}_2, \hat{\bd{m}} \}$ 
shown in Fig.~\ref{fig:tilt}.
Introducing spherical basis vectors $\bd{e}^p =  \bd{e}_1 + i p \bd{e}_2$  with $p = \pm$,
we write $\bd{S}_i  =  S^{\parallel}_i \hat{\bd{m}} + \bd{S}_i^\bot$ and
$ \bd{S}_i^{\bot}  =  \frac{1}{2} \sum_{ p = \pm } S^{-p}_i \bd{e}^p$.
We then express the spin components in terms of canonical boson operators $b_i$ using the
Holstein-Primakoff transformation, 
$ S_i^{\parallel}  =  S -  b^{\dagger}_i b_i$,
$ S_i^+  =( S_i^{-} )^{\dagger} =    [ 2S  - b^{\dagger}_i b_i ]^{1/2}   b_i $.
Retaining only quadratic terms in the bosons
and Fourier transforming,
$ b_i = \frac{1}{\sqrt{N}} \sum_{ {k}} e^{ i k x_i } b_{k}$,
our bosonized spin 
Hamiltonian is approximated by
\begin{equation}
 {\cal{H}} \approx {\cal{H}}_0 + \sum_{{k}}
 \Bigl[ A_{{k}} b^{\dagger}_{{k}} b_{{k}} + \frac{ B_{{k}}}{2} \bigl( 
  b^{\dagger}_{ {k}} b^{\dagger}_{ - {k}} +b_{ -{k}} b_{  {k}}
   \bigr) \Bigr],
 \label{eq:H2}
 \end{equation}
with 
 \begin{eqnarray}
A_k & = &  
\Bigl[ \frac{D_0}{3} + \frac{D_k}{6} \Bigr] ( 3 \sin^2 \vartheta -1 ) + J_0 - J_k 
 + h \cos \vartheta    \hspace{7mm}
 \end{eqnarray}
 and $B_k  =      - \frac{D_k }{2} \cos^2 \vartheta $, where
$J_k =  2 J S \cos ( k a )$ and 
\begin{equation}
 D_k =  \frac{D_0}{\zeta (3 )} 
\sum_{ n=1}^{\infty} \frac{ \cos ( n k a )}{n^3}.
 \end{equation} 
The infinite series
 $
\sum_{ n=1}^{\infty} \frac{ \cos ( n k a )}{n^3}$ 
 represents the so-called Clausen function
 ${\rm Cl}_3 ( ka )  =
 {\rm Re} \,  {\rm Li}_3 ( e^{ ika} )$, where
${\rm Li}_3 ( z )$  is the polylogarithm.\cite{Lewin81}
From the known series expansion of  ${\rm Li}_3 ( e^{ \mu} )$ 
we obtain for  $|k a| \ll  1$,
 \begin{equation}
 D_k / D_0 =  1   -   \frac{3}{2}   (k a)^2  d_1 \ln (d_2/| ka |)  + {\cal{O}} ( (ka)^3 )  ,
 \label{eq:Dkasym}
 \end{equation}
where 
 $d_1 = 1/ [3 \zeta ( 3)]  $ and $d_2 = e^{3/2}$.
Using a Bogoliubov transformation to diagonalize
the Hamiltonian (\ref{eq:H2}) we obtain the
magnon dispersion $E_{{k}} = \sqrt{ A_{{k}}^2 - | B_{{k} } |^2} $,
which is shown graphically in Fig.~\ref{fig:dispersion}.
For simplicity we set $J=0$ from now on.
%
%
For $| ka | \ll 1$ and $ | h - D_0 | \ll D_0$
the magnon dispersion is then approximated by
 $ E_k \approx \sqrt{ \Delta^2 + v_k^2 k^2 }$,
where the square of the gap is 
$ \Delta^2 = D_0 ( h - D_0)$ for $h > D_0$, and
$\Delta^2 = 2 D_0 ( D_0 -h )$ for $ h < D_0$.
The squared velocity $v_k^2$ exhibits a logarithmic divergence
for small wave vectors,
$ v_k^2  =  
  v_0^2 \ln (1 / | ka | ) + c_0^2 $,
with 
 $v_0^2  =  (D_0 a)^2 d_1$ and 
 $ c_0^2  = \frac{3}{2}  v_0^2$. 
The logarithmic divergence of $v_k$ 
is a unique signature of the $1/|x|^3$-decay of the
dipolar interaction in one dimension.
Indeed, logarithmic corrections to the
dispersion are characteristic of one-dimensional systems
with long-range interactions decaying as
$1/|x|^\beta$ with odd $\beta$ (see Refs.~\onlinecite{Li91} and \onlinecite{Schulz93} for the case of the
Coulomb interaction, $\beta = 1$, and Ref.~\onlinecite{Inoue06} for 
the case of an arbitrary $\beta$).

\begin{figure}[tb]
  \centering
  \includegraphics[keepaspectratio,width=\linewidth]{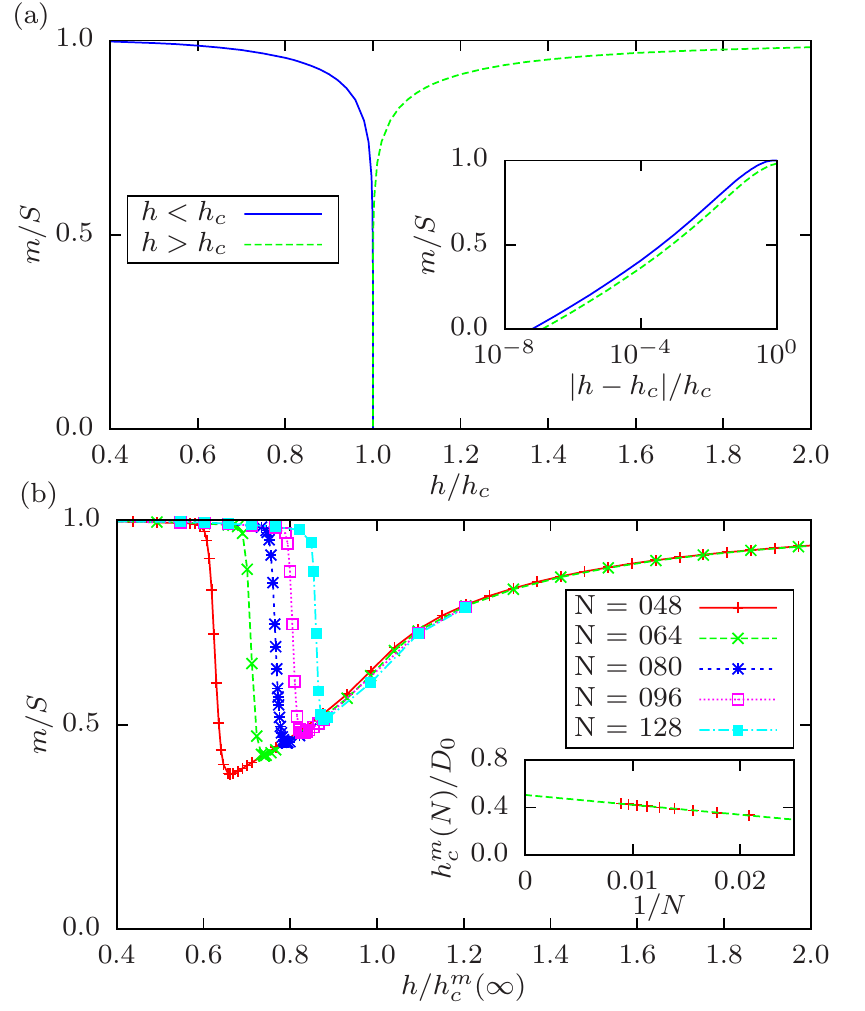}
  \caption{(Color online)
(a) Spin-wave result for the one-loop corrected total magnetic moment $m$ given
in Eq.~(\ref{eq:mres}) as a function of $h/h_c$ for $S\! =\! 1/2$. The curves
are for $J\!=\!0$ and $h_c\! =\! D_0$. The inset shows $m/S$ versus the logarithm
of the reduced magnetic field $\vert h - h_c \vert /h_c$ for $h\! <\! h_c$
(solid line) and for $h\! >\! h_c$ (dashed line). 
(b) Corresponding DMRG results. The critical magnetic field $h_c^m(\infty)\!
\approx\! 0.507 D_0$ was determined by extrapolation  of the critical fields
$h_c^m(N)$ of $N$-site chains with open boundary conditions, as shown in the inset.
}
  \label{fig:magnetization}
\end{figure}

Despite the unusual logarithmic correction to the dispersion
in Gaussian approximation, 
it seems at first glance that
our spin-wave approach remains valid
for all values of the magnetic field.
This is not the case, however, because
in a narrow range of magnetic fields close to $D_0$
the leading quantum correction to the
magnetic moment $m$ per site
completely overwhelms the classical
result $m \approx S$.
Retaining the leading $1/S$ quantum correction to $m$, 
we obtain
 \begin{equation}
 m= \frac{1}{N} \sum_i \langle S_i^{\parallel} \rangle =
S + \frac{1}{2} - \frac{1}{N} \sum_k \frac{ A_k}{2 E_k}.
 \label{eq:mres}
 \end{equation}
Because $E_{k=0} $ vanishes for $h \rightarrow D_0$ while
$A_{k=0}$ remains finite for sufficiently small $| h-D_0|$, the 
last term in Eq.~(\ref{eq:mres})
becomes arbitrarily large, which is also evident from the 
numerical evaluation of Eq.~(\ref{eq:mres}) shown
in Fig.~\ref{fig:magnetization}~(a).

Obviously, for any $S < \infty$ there is a range of magnetic fields 
where the tilt angle is small and the leading quantum correction to the
magnetization is larger than the classical result $m \approx S$. In this regime
our simple spin-wave approach breaks down and we need more sophisticated methods
to investigate the behavior of the system. We have studied the quantum critical
regime $ | h -h_c | \lesssim h_c$ using both the numerical density-matrix
renormalization group (DMRG)\cite{DMRGref} and the analytical functional
renormalization group (FRG)\cite{Kopietz10}  methods.

\section{DMRG approach}

Using the DMRG, we have calculated the ground state and its magnetization for
systems of up to $N=128$ sites and a varying bulk magnetic field $h$ in the
$z$-direction. In the simulations, we kept up to 320 density-matrix eigenvalues,
leading to a maximum discarded weight of $10^{-12}$. It should be noted that
DMRG simulations of the dipolar spin chains are numerically demanding for
several reasons. 
First, the long-range nature of the interaction leads to increased correlations
between different parts of the system that have to be encoded in the variational
ground state. Hence, the number of states that need to be kept is rather high
compared to the usual spin chains with nearest-neighbor Heisenberg interaction.
Second, the $SU(2)$ spin symmetry is broken, so that $S_z$ is not a good quantum
number. Therefore, the restriction of the basis states to a certain spin
component $S^z$, which is normally used to significantly increase the efficiency
of the DMRG, is not possible. Finally, for $h\!<\!h_c$ the ground state is
two-fold degenerate due to the $Z_2$ symmetry $S^x_i\!\rightarrow\!-S_i^x$.
Accordingly, it is crucial that we mix two states into the density matrix and
target both states of the ground-state doublet. In the $h/h_c\!\rightarrow\! 0$
limit, the system is fully polarized in either the $+x$- or the $-x$-direction,
but while the $Z_2$ symmetry is spontaneously broken in the infinite chain, the
variational DMRG ground state, which always describes a finite system,
is an arbitrary linear combination of two states with opposite polarization
directions. Since the mixing angle is not fixed, the measured magnetization in
the $x$-direction is random. We have found that we can obtain reliable results
by adding a small local magnetic field $h_x=10^{-10} \mu^2/a^3$ on the end
sites. This field explicitly breaks the $Z_2$ symmetry and, consequently, leads
to a unique ground state while generating an energy difference of  the order
$10^{-10} \mu^2 / a^3$ in the formerly degenerate doublet states.

In finite systems, the correlation length $\xi$ is bounded by the system size
$L$. Therefore, phase transitions are strictly possible only in infinite
systems; the inverse size $1/L$ plays the role of an additional parameter that
moves the system away from the critical point. To determine the critical
parameters in the thermodynamic limit, the dependence of thermodynamic
quantities on the system size can be investigated with finite-size scaling
theory. 
In this context, one uses the notion of a pseudo-critical field $h_c(L)$
associated with indications of critical behavior in a finite system of
size $L$. 
Note that, for a given system, multiple definitions of
$h_c(L)$ that result in the correct value of $h_c$ in the
limit $L\rightarrow \infty$ are possible.
In general, one finds that the pseudo-critical
field $h_c(L)$ scales as $L^{-{1}/{\nu}}$, where $\nu$ is the correlation-length
critical exponent and $\xi\!\propto\! \vert h\! -\! h_c \vert^{-\nu}$.
The order parameter scales as $m_x(h_c,L)\! \propto\! L^{-{\beta}/{\nu}}$,
with $\beta$ the order-parameter critical exponent. 
For the two-dimensional Ising universality
class, the relevant critical exponents have the values $\nu$=$1$ and
$\beta$=$1/8$.

In Fig.~\ref{fig:magnetization}~(b) we show our numerical result for the total
magnetic moment
 \begin{equation}
 m = \sqrt{m_x^2 + m_y^2 + m_z^2} \,,  \qquad 
 m_\alpha = \frac{1}{N} \sum_{i} \langle S_i^\alpha \rangle \,. 
 \end{equation}
Since $m_y$ vanishes for the
field direction selected here, the total magnetic moment is determined 
by $m_x$ and $m_z$ only. 
While the value of $m_x$ drops from about one half to zero in
the vicinity of the pseudo-critical field $h_c(L)$, $m_z$ increases linearly for
$h \lesssim h_c(L)$ and enters a saturation regime for $h \gtrsim h_c(L)$. Consequently, $m$ exhibits a minimum close to $h_c(L)$. By determining the position $h_c^m(N)$ of this minimum for systems of different numbers of sites $N$=$L/a$, we obtain an approximate value
$h_c^m(\infty) \! \approx \! 0.507 D_0$ upon extrapolating to $N \rightarrow
\infty$.

In contrast to the spin-wave results in
Fig.~\ref{fig:magnetization} (a), the total moment is finite for any $h$ and
exhibits a large asymmetry, indicating that the Ginzburg regime where non-Gaussian
fluctuations are important is very broad on the disordered side
$(h > h_c)$ of the transition.
This corresponds to a regime where the order parameter $m_x$ is zero
but the magnetization $m_z$ along the field direction is not yet saturated
due to large correlations $\langle S_i^x S_j^x \rangle$ in the $x$-component of the spin.

\begin{figure}[tb]    
  \centering
  \includegraphics[keepaspectratio,width=\linewidth]{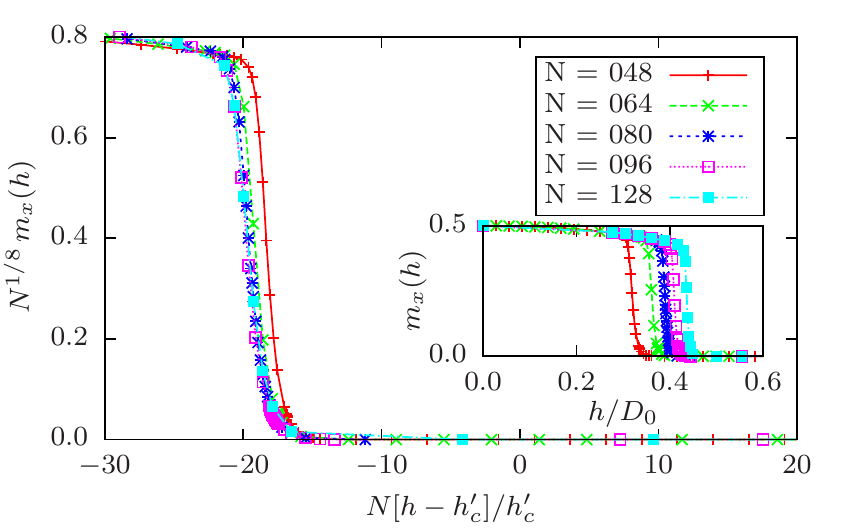}
  \caption{\label{fig:4}(Color online) %
DMRG results for the order parameter $m_x = \frac{1}{N} \sum_i \langle S^x_i
\rangle$ as a function of $[h-h_c']/h_c'$ for $S\!=\!1/2$ and different $N$.
The main plot shows the scaled magnetization $N^{{\beta}/{\nu}} m_x$
with exponents $\beta\! =\! {1}/{8}$ and $\nu \!=\!1$
of the two-dimensional Ising model versus $N [h-h_c']/h_c'$, with
$h_c'\!=\! 0.516 D_0$ chosen to optimize the collapse of the data.
The inset shows the raw data as a function of $h/D_0$.
} \end{figure}

In order to ascertain whether the phase transition belongs to the Ising
universality class, we also investigate the behavior of the order
parameter $m_x$ separately.
In Fig.~\ref{fig:4}, we plot the scaled magnetization $N^{\beta/\nu}
m_x(h,N)$ with $\nu$=$1$ and $\beta$=$1/8$ versus $N [h - h_c'] /h_c'$ for
different values of $N$=$L/a$, where the critical field $h_c'$ is 
chosen to optimize the collapse of the data.
The collapse of the data onto a single curve for a 
critical field $h_c' \approx 0.516 D_0$ shows that our numerical
results for the magnetization 
are consistent with the two-dimensional Ising universality class. 
The value $h'_c$ estimated from the best collapse of the
data and the value $h_c^m(\infty)$ obtained by extrapolating the minimum
of $m/S$ in Fig.~\ref{fig:magnetization} (b) differ by only $2.1\%$.
We ascribe the imperfect collapse of the $N$=$48$ data to the fact that
the dipolar interaction is long-range 
so that sufficiently large systems have to be treated before 
the true critical behavior manifests itself.

\section{FRG approach}

To investigate our model for general $S$ and to verify that
the RG flow  in the vicinity of the critical point
indeed is consistent
with the Ising universality class, we have 
studied the effect of spin-wave interactions 
using the functional renormalization group.\cite{Kopietz10}
Since we are interested in the critical fluctuations,
we may simplify the calculations using the
hermitian field parametrization of the spin-wave interactions\cite{Hasselmann06}
in which 
we express the Holstein-Primakoff bosons $b_k$
in terms of two canonically conjugate hermitian field operators,
 $b_{{k}}  =   [    \phi_{{k}} \sqrt{h_k}   + i \Pi_{{k}}/ \sqrt{h_k}]/ \sqrt{2}$,
where $h_k  =  A_k + | B_k |$. 
Since the fluctuations of the canonical momentum $\Pi_{{k}}$ remain gapped
at the quantum critical point,
we may integrate over the field  $\Pi_{{k}}$ in Gaussian approximation and obtain an effective
Euclidean action $S_{\rm eff} [ \phi ]$ for the field $\phi_k$
describing the critical fluctuations.
The Gaussian propagator of the $\phi$-field is then
$G_0 ( K ) = [ \omega^2 + E_k^2]^{-1}$, where  
$K = (k , i \omega )$ denotes both momentum and frequency,
and the
effective action $S_{\rm eff} [ \phi ]$ of the critical
fluctuations can be expanded as
 \begin{eqnarray}
 S_{\rm eff} [ \phi ] & = &  \frac{1}{2T} \sum_K G_0^{-1} ( K )
 \phi_{-K} \phi_K + \frac{1}{T}
\sum_{n=0, n \neq 2 }^{\infty}    \frac{1}{n!  N^{\frac{n}{2} -1}}
 \nonumber
 \\
 & & \hspace{-13mm} \times 
\sum_{ K_1  \ldots  K_n }
 \delta_{ K_1+ \cdots + K_n,0 } 
\Gamma^{(n)}_0 ( k_1 \ldots k_n  ) \phi_{K_1} \cdots \phi_{ K_n},
 \label{eq:Seff} 
\end{eqnarray}
with $T$ the temperature. The interaction vertices $\Gamma^{(n)}_0 ( k_1 \ldots k_n  )$ 
can be expressed in terms of the Fourier transform $D_k$ of the dipole-dipole interaction by expanding the spin operators in powers of the Holstein-Primakoff bosons $b_k$ and
then setting  $b_{{k}}  \rightarrow     \phi_{{k}} \sqrt{h_k / 2}$.
For vanishing external momenta,
the interaction  vertices have finite limits
 $\Gamma^{(n)}_0 \propto S^{ 1 - \frac{n}{2}}$, and
in the symmetric phase ($ h > h_c$) the odd vertices vanish.

The magnon spectrum
can be obtained from the poles of the true
propagator
$G ( K ) = [ G_0^{-1} ( K ) + \Sigma ( K ) ]^{-1}$,
where $\Sigma ( K )$ is the irreducible self-energy
of the effective field theory defined in Eq.~(\ref{eq:Seff}).
Since 
the higher order vertices involve increasing powers
of $1/S$ for large $S$, one could 
calculate $\Sigma ( K )$
perturbatively.
However, the  corrections become arbitrarily large 
for $h \rightarrow h_c$, which is not surprising
because all higher order vertices of our $1+1$-dimensional field 
theory 
are relevant at the Gaussian fixed point,
with canonical dimension $+2$.
To regulate the infrared singularities, we replace
the inverse Gaussian propagator by 
 $G_{0, \Lambda}^{-1} ( K ) = G_0^{-1} ( K ) + R_{\Lambda} ( k )$,
using the regulator function $R_{\Lambda} ( k )$
proposed by Litim.\cite{Litim01}
In the limit of vanishing flow parameter
$\Lambda$, the regulator vanishes, so that we recover
our original model.

\begin{figure}[tb]    
  \centering
  \includegraphics[keepaspectratio,width=.7\linewidth]{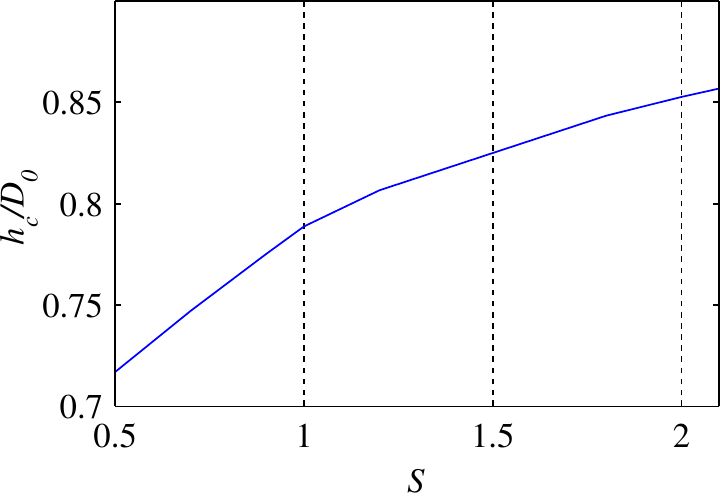}
  \caption{\label{fig:5}(Color online) %
FRG results for the critical field $h_c^{(S)}$ as a function of the spin $S$.
For the relevant values of $S>1/2$ that might be investigated in a future DMRG study
we find $h_c^{(1)} = {0.79 D_0}$, $h_c^{(3/2)} = {0.83 D_0}$, and $h_c^{( 2 )} = {0.85 D_0}$.
} \end{figure}

We use 
a simple truncation
of the formally exact hierarchy of FRG flow 
equations\cite{Kopietz10} for scalar field theories of the type
(\ref{eq:Seff}) to re-sum the perturbation series,
expanding 
the flowing self-energy as
\begin{equation}
  \Sigma_{\Lambda} ( K ) = \Sigma_{\Lambda} (0)
  + Y_{\Lambda} k^2 +  ( Z_{\Lambda}^{-1} -1 ) \omega^2 
  +  {\cal{O}} ( k^3 , \omega^3),
 \label{eq:selfexpand}
 \end{equation}
with flowing coupling constants $\Sigma_{\Lambda} (0)$,
$Y_{\Lambda}$, and $Z_{\Lambda}$.
The scale-dependent long-wavelength magnon spectrum is then
 \begin{equation} 
  {E}_{\Lambda , k }^2  =  \Delta_{\Lambda}^2 +
  \left[ v_{\Lambda}^2   \ln (1/ | k a | ) + c_{\Lambda}^2 \right] k^2, 
 \end{equation}
where the renormalized squared gap is
$ \Delta_{\Lambda}^2  =  Z_{\Lambda} [ \Delta^2 + \Sigma_{\Lambda} ( 0 ) ]$,
and the renormalized magnon velocities are given by
$ v_{\Lambda}^2  =  Z_{\Lambda}  v_0^2 $ and
$ c_{\Lambda}^2  =  Z_{\Lambda}  [ c_0^2 + Y_{\Lambda} ]$.
Note that 
in general the low energy expansion of the self-energy
can also contain terms proportional to $k^2 \ln (1/| k a | )$ and $\omega^2 \ln(1/ | k a |)$,
which are of the same order
as the terms retained in Eq.~(\ref{eq:selfexpand}).
However, we find that the flowing self-energy $\Sigma_{\Lambda} ( K )$ is
analytic in $K=0$ for any finite $\Lambda$, due to the presence of the
regulator function $R_{\Lambda} ( k )$, so that  
these terms do not appear in our FRG approach.
Moreover, in our truncation we retain only the
momentum- and frequency-independent parts $\Gamma^{(3)}_{\Lambda}$ and
$\Gamma^{(4)}_{\Lambda}$ of the three-point and four-point vertices.
This truncation is then not sufficient to 
calculate the critical exponent $\eta$ for the anomalous dimension, which is determined
by the frequency-dependent part of the
four-point vertex.

Analyzing the structure of the FRG flow equations in the vicinity of
the quantum critical point, corresponding to a non-Gaussian 
fixed point of the flow equations,\cite{Fisher74Wilson75} 
we find that the FRG confirms the Ising 
universality class expected from the general symmetry arguments and 
numerically derived within the DMRG analysis. 
For $h=h_c$ the wave function renormalization factor $Z_{\Lambda}$ 
and the rescaled three-point vertex 
$\gamma^{(3)}_{\Lambda} \propto Z_{\Lambda}^{3/2} \Gamma^{(3)}_{\Lambda}/(\Lambda a)^2$ 
flow to zero in the limit $\Lambda \to 0$, while 
the rescaled four-point vertex 
$\gamma^{(4)}_{\Lambda} \propto Z_{\Lambda}^{2} \Gamma^{(4)}_{\Lambda}/(\Lambda a)^2$ 
approaches a finite fixed-point value.
In particular, the three-point vertex 
is marginally irrelevant at the fixed point, flowing
asymptotically as $\gamma^{(3)}_{\Lambda} \sim [\ln(\Lambda_0/\Lambda)]^{-1/2} $
in the limit $\Lambda \to 0$. 
(Here $\Lambda_0 \sim 1/a$ is the initial RG scale,
corresponding to an ultraviolet cutoff).
Importantly, we also find that the logarithmic correction to
the spin-wave velocity becomes marginally irrelevant at the fixed point,
namely $v_{\Lambda}^2 \sim [\ln(\Lambda_0/\Lambda)]^{-r} $ for $\Lambda \to 0$,
with $r \approx 0.017$, while $c_{\Lambda}^2$ remains finite.
This signals that scale invariance is eventually restored at the critical point,
strengthening the confidence in the finite-size scaling analysis of the DMRG results.

Besides investigating the behavior of the RG flow in the vicinity of the non-Gaussian fixed point,
we may also use the FRG approach to estimate the critical field $h_c^{(S)}$ for different values
of the spin $S$.
For a given magnetic field $h<h_c$, we fine-tune the tilt angle $\vartheta(h)$ so that the linear vertex
$\Gamma^{(1)}_{\Lambda}$ vanishes for $\Lambda \rightarrow 0$ and fix the value of the
critical field $h_c$ where the magnetization $m$ drops more steeply, corresponding to the limit $\vartheta(h_c) \to 0$.
For $S=1/2$ we obtain $h_c^{(1/2)} \approx{0.72 D_0}$,
which is significantly smaller than the classical result $h_c = D_0$. However,
the DMRG result $h_c \approx 0.51 D_0$ is even smaller, showing that 
our FRG truncation probably still misses some sizable effects of
quantum fluctuations.
In Fig.~\ref{fig:5} we plot our FRG results for higher values of the spin.

\section{Conclusions}

In this work, we have
calculated the magnon spectrum and the magnetization curve of
dipolar spin chains in a transverse magnetic field using spin-wave theory and
renormalization group methods.
We have shown that at a critical field $h = h_c$ the system exhibits a quantum critical point
belonging to the two-dimensional Ising universality class.
Outside the critical regime,
where the excitation spectrum is well described by linear spin-wave theory,
the magnon velocity
exhibits a logarithmic dependence on the wave vector, which 
might be useful to characterize
dipolar interactions in experimental realizations of spin chains.
Using the numerical density-matrix renormalization group method,
we have presented quantitatively accurate results for the magnetization curve
and the location of the critical point for the case of spin $S=1/2$, finding a strong 
reduction in the value of $h_c$ in comparison to the classical limit ($S \to \infty$). 
Finally, analyzing an effective low-energy field theory for our model 
by means of the functional renormalization group method, we 
have pointed out the emergence of a
scale-invariant excitation spectrum in the vicinity of the critical point,
where the logarithmic correction to the magnon velocity becomes an irrelevant perturbation,
thus confirming the consistency of our numerical DMRG results with the expected 
universality class.

\begin{acknowledgments} 

This work was financially supported by the DFG via SFB/TRR49 and FOR 723.

\end{acknowledgments}

\end{document}